\documentclass[layout=twocolumn,journal=aesccq,manuscript=article]{achemso}
 \setkeys{acs}{articletitle = true}
 \usepackage{natbib}

\usepackage{graphicx,color}
\usepackage{amsmath,amsfonts,amssymb}
\usepackage{dcolumn}

\newcommand{\fig}{Fig.}

\newcommand{\figref}[1]{\fig~\ref{#1}}

\newcommand{\tabref}[1]{table~\ref{#1}}

\renewcommand{\eqref}[1]{equation~(\ref{#1})}

\newcommand{\rxnref}[1]{reaction~(\ref{#1})}

\usepackage{listings}

\usepackage{tabularx}
\usepackage{extarrows}

\newcounter{defcounter}
\newenvironment{reaction}{%
\addtocounter{equation}{-1}
\refstepcounter{defcounter}

\begin{equation}}
{\end{equation}}

\let\oldmaketitle\maketitle
\let\maketitle\relax

\makeatletter
\def\@dotsep{4.5}
\makeatother
\title{Atom Tunneling in the Water Formation Reaction H$_2$~+~OH~$\rightarrow$~H$_2$O~+~H on an Ice Surface}
\author{Jan Meisner}
\email{meisner@theochem.uni-stuttgart.de}
\author{Thanja Lamberts}
\author{Johannes K\"{a}stner}
\affiliation{Institute for Theoretical Chemistry, University of Stuttgart, Pfaffenwaldring 55, 70569 Stuttgart,Germany}

\date{}

\begin{document}

\twocolumn[
\begin{@twocolumnfalse}
\oldmaketitle

\begin{center}
  \date{\today}
\end{center}

\begin{abstract}
OH radicals play a key role as an intermediate in the water formation chemistry of the  interstellar medium. For example the reaction of OH radicals with H$_2$ molecules is among the final steps in the astrochemical reaction network starting from O, O$_2$, and O$_3$. Experimentally it was shown that even at 10\,K this reaction occurs on ice surfaces. 
As the reaction has a high activation energy only atom tunneling can explain such experimental findings. 

In this study we calculated reaction rate constants for the title reaction on a water-ice I$_h$ surface. To our knowledge, low-temperature rate constants on a surface are not available in the literature. All surface calculations were done using a QM/MM framework (BHLYP/TIP3P) after a thorough benchmark of different density functionals and basis sets to highly accurate  correlation methods. Reaction rate constants are obtained using  instanton theory which takes atom tunneling into account inherently, with constants down to 110\,K for the Eley--Rideal mechanism and down to 60\,K for the Langmuir--Hinshelwood mechanism. We found that the reaction is nearly temperature independent  below 80\,K. 
We give kinetic isotope effects for all possible deuteration patterns for both reaction mechanisms.  
For the implementation in astrochemical networks, we also give fit parameters to a modified Arrhenius equation. Finally, several different binding sites and binding energies of OH radicals on the I$_h$ surface are discussed and the corresponding rate constants are compared to the gas-phase case.

\end{abstract}
\end{@twocolumnfalse}
]



\maketitle

\section{Introduction}

Water ice was first detected in 1973 \cite{Gillett:1973} and is meanwhile known to be the main component of most interstellar ices. \cite{Dishoeck:2004,Oberg:2011,boo15} Therefore, the surface formation of water in space was studied extensively experimentally in ultra-high vacuum setups \cite{hiraoka1998, miyauchi2008, oba2009, Ioppolo:2008, ioppolo2010, cuppen2010B, rom11c, oba2014, lamberts14, Lamberts:2016, Hama:2016} and through modeling studies with different varieties of Kinetic Monte Carlo \cite{cup07, Chang:2007, Cazaux:2010, Garrod:2013, Vasyunin:2013, lam14b} and rate equation models. \cite{fur13,taq13,furuya2017} Although water can also be formed via gas-phase reactions, it is the formation on the surface of dust grains in dense molecular clouds that can explain the observed abundances. One of the crucial factors is that the surface provides an efficient way for the reaction products to lose their excess energy. \cite{buk15} In other words, addition reactions which yield only one reaction product cannot take place in the gas phase, but can take place on the surface. For more insight on the gas-phase routes we refer the reader to the recent review by Van Dishoeck \emph{et al.} \cite{dis13} and focus on surface chemistry from hereon. An involved network of surface reactions in the interstellar medium (ISM) was originally proposed by Tielens and Hagen \cite{tie82} and has been updated incorporating the experimental results mentioned above.\cite{lamberts14,lam14b} Three main water formation routes constitute this network: hydrogenation of atomic oxygen, molecular oxygen (O$_2$), and ozone (O$_3$). Depending on the interstellar region of interest different routes dominate. In each pathway the hydroxyl radical is formed and subsequently reacts with either atomic or molecular hydrogen.

The barrierless direct hydrogenation of O atoms by H atoms is believed
to be important in translucent and diffuse clouds in which H atoms are
more abundant than H$_2$ molecules\cite{cup07} and was experimentally studied by different groups.\cite{hiraoka1998,dul10,jin11}
\begin{reaction}
  \text{O}  + \text{H}    \rightarrow  \text{OH}
\end{reaction}

Alternatively, O$_2$ can be hydrogenated twice and the resulting hydrogen peroxide (H$_2$O$_2$) reacts with another H atom to form water and a hydroxyl radical:
\begin{reaction}
  \text{O}_2 + \text{H}  \rightarrow \text{H} \text{O}_2
  \label{rxn:h3o:o2+h}
\end{reaction}
\begin{reaction}
  \text{H}  \text{O}_2 + \text{H}  \rightarrow \text{H}_2 \text{O}_2 
  \label{rxn:h3o:ho2+h}
\end{reaction}
\begin{reaction}
  \text{H}_2 \text{O}_2 + \text{H}  \rightarrow \text{H}_2 \text{O} + \text{OH} 
  \label{rxn:h3o:h2oh+h}
\end{reaction}
These reactions have been shown to proceed even at temperatures as low as 12\,K.\cite{miyauchi2008,oba2009,cuppen2010B} The last reaction again proceeds via a barrier and has been shown to take place via tunneling both experimentally \cite{oba2014} and using instanton theory. \cite{lam16} Note that the sequential hydrogenation of O$_2$ up to H$_2$O$_2$ is not possible in the gas phase, because of the reasons mentioned above. The reaction between H and HO$_2$ can also result in two OH radicals via a decomposition of activated H$_2$O$_2^*$ \cite{key86,mou07}
\begin{reaction}
  \text{H}_2 \text{O}_2^* \rightarrow 2 \, \text{OH} 
  \label{rxn:h3o:2ohformation}
\end{reaction}
The resulting OH radicals can recombine to H$_2$O$_2$ or form water and an O atom:
\begin{reaction}
  2 \, \text{OH}   \rightarrow  \text{H}_2\text{O} + \text{O} 
  \label{rxn:h3o:2oh}
\end{reaction}
Microscopic kinetic Monte Carlo modeling found that experiments can be best described by the sequence 
\ref{rxn:h3o:o2+h}, \ref{rxn:h3o:ho2+h} leading to H$_2$O$_2^*$ and \ref{rxn:h3o:2ohformation} with OH radical recombination to H$_2$O$_2$ dominating over reaction \ref{rxn:h3o:2oh}. \cite{lamberts2013, lam14b}

Finally, O$_3$ can be hydrogenated which leads to an O$_2$ molecule and OH radical and has been experimentally studied in the solid state at 10\,K. \cite{mok09,rom11c}
\begin{reaction}
  \text{O}_3 + \text{H}  \rightarrow \text{O}_2 + \text{OH}  
  \label{rxn:h3o:ozone} 
\end{reaction}

In all of these reaction pathways, OH radicals are formed. Subsequent reaction to form water can take place via reaction \ref{rxn:h3o:OH+H} or \ref{rxn:h3o:h2+oh}:
\begin{reaction}
  \text{H}   + \text{OH}   \rightarrow  \text{H}_2\text{O} 
  \label{rxn:h3o:OH+H}
\end{reaction}
\begin{reaction}
  \text{H}_2 + \text{OH}  \rightarrow \text{H}_2\text{O} + \text{H}\;.
  \label{rxn:h3o:h2+oh}
\end{reaction}
Reaction \ref{rxn:h3o:OH+H} is barrierless, since it is a radical recombination reaction, but the reaction with H$_2$ proceeds via a barrier. It is the topic of this paper. A hydrogen atom is transferred from the H$_2$ molecule to the OH radical to finally form water. In the ISM the competition between reactions \ref{rxn:h3o:OH+H} and \ref{rxn:h3o:h2+oh} is determined by the interstellar environment, the relative abundances of H, H$_2$, and O in the gas phase, and the reaction rates. Modeling this process in dense molecular clouds -- where ice layers are thick and the H$_2$ abundance is high -- therefore requires detailed knowledge of the low-temperature reaction rate constant. 
Reaction (\ref{rxn:h3o:h2+oh}) in the gas phase was studied extensively experimentally \cite{rav81,tal96,orkin2006} and computationally.\cite{matzkies1998,manthe2000,ngu10,nguyen2011,mei16a} The high activation energy of 2000\,K (experimentally determined by laser-induced fluorescence after photolysis)\cite{tal96,atk04} to  3000\,K (computationally determined) \cite{nguyen2011,mei16a} shows that atom tunneling is crucial for the reaction rate at low temperatures.\cite{manthe2000,nguyen2011,mei16a} Oba \emph{et al.} found experimental evidence that the reaction of H$_2$ and OH to water and hydrogen atoms also occurs on surfaces even at 10\,K due to atom tunneling. \cite{oba2012} Recently, we published an extensive study on the reaction of molecular hydrogen and hydroxyl radicals (\eqref{rxn:h3o:h2+oh}) in the gas phase including all possible isotope patterns down to 80\,K.\cite{mei16a} Here, we extend this work to the adsorption of OH radicals onto crystalline I$_h$ water ice and the subsequent reaction with H$_2$.
We present binding sites, energies, reaction paths and reaction rate constants for the reaction of H$_2$ with OH on the surface. This includes the effect of atom tunneling at low temperatures via the use of instanton theory. 
\subsection*{}

This paper is structured as follows:
In the methods section we present a benchmark to find a suitable density functional and basis set
by comparing the quality with highly correlated calculations on UCCSD(T)-F12\cite{adl07,adl09} level.
Furthermore, the water ice surface and the organization of the QM/MM setup,
as well as the methodology of the reaction rate calculations are described.
In the results section we show binding sites and binding energies of the OH radical on the ice surface as well as accompanying activation energies, transition state structures and intrinsic reaction coordinates (IRC).
We give reaction rate constants for the Eley--Rideal and the Langmuir--Hinshelwood mechanism using multidimensional atom tunneling via semiclassical instanton theory \cite{lan67,mil75,col77,cal77,gil77} and make a comparison to the analytical solution of an Eckart shaped barrier. 
The results are compared to those in the gas phase and the impact of surfaces on the 
reactivity is discussed. 
Kinetic isotope effects for all eight possible permutations of exchanging hydrogen for deuterium are shown as well. 
The last section discusses the implications to astrochemistry, gives fits of to a modified 
Arrhenius equation and concludes the study.

\section{Methods}

\subsection{Choice of Electronic Potential}

In order to obtain reliable reaction rate constants,
the method to calculate the underlying electronic potential has to be as accurate as possible.
Instanton calculations using highly correlated wave function methods 
are too time consuming so we applied density functional theory (DFT) throughout this work.  
For this purpose, an extensive benchmark of the most common functionals and basis sets 
has been performed, 
as can be seen in the Supplementary Information.
The BHLYP functional \cite{dir29,sla51,bec88,lee88,becke1993a} in combination 
with the def2-SVPD basis set\cite{rap10} 
is found to describe the reaction well.
The BHLYP functional has previously been found to describe astrochemical reactions 
with open-shell molecules properly.\cite{andersson2004,rim14}
It was also found appropriate for water-water interactions
(see Supplementary Information). Thus BHLYP is used in the remainder of this paper.
All DFT calculations were performed with Turbomole version 7.0.\cite{turbomole}
SCF energies were converged to an accuracy of $10^{-9}$ Hartree on an
\emph{m5} grid.\cite{eichkorn1997}

To include the environment, \emph{i.e.}, the water surface,
we used a QM/MM framework for which we used the ChemShell interface.\cite{she03,met14}
The QM part can be polarized by means of electrostatic embedding into the MM charges.
All geometry optimizations and reaction rate calculations were done with DL-FIND.\cite{kae09a}
The visualisation of the molecules and ice surface was done using Visual Molecules Dynamics (VMD).\cite{humphrey1996}

\subsection{Surface Model and QM/MM setup}

\begin{figure}[h!]
\begin{center}
  \includegraphics[width=8cm]{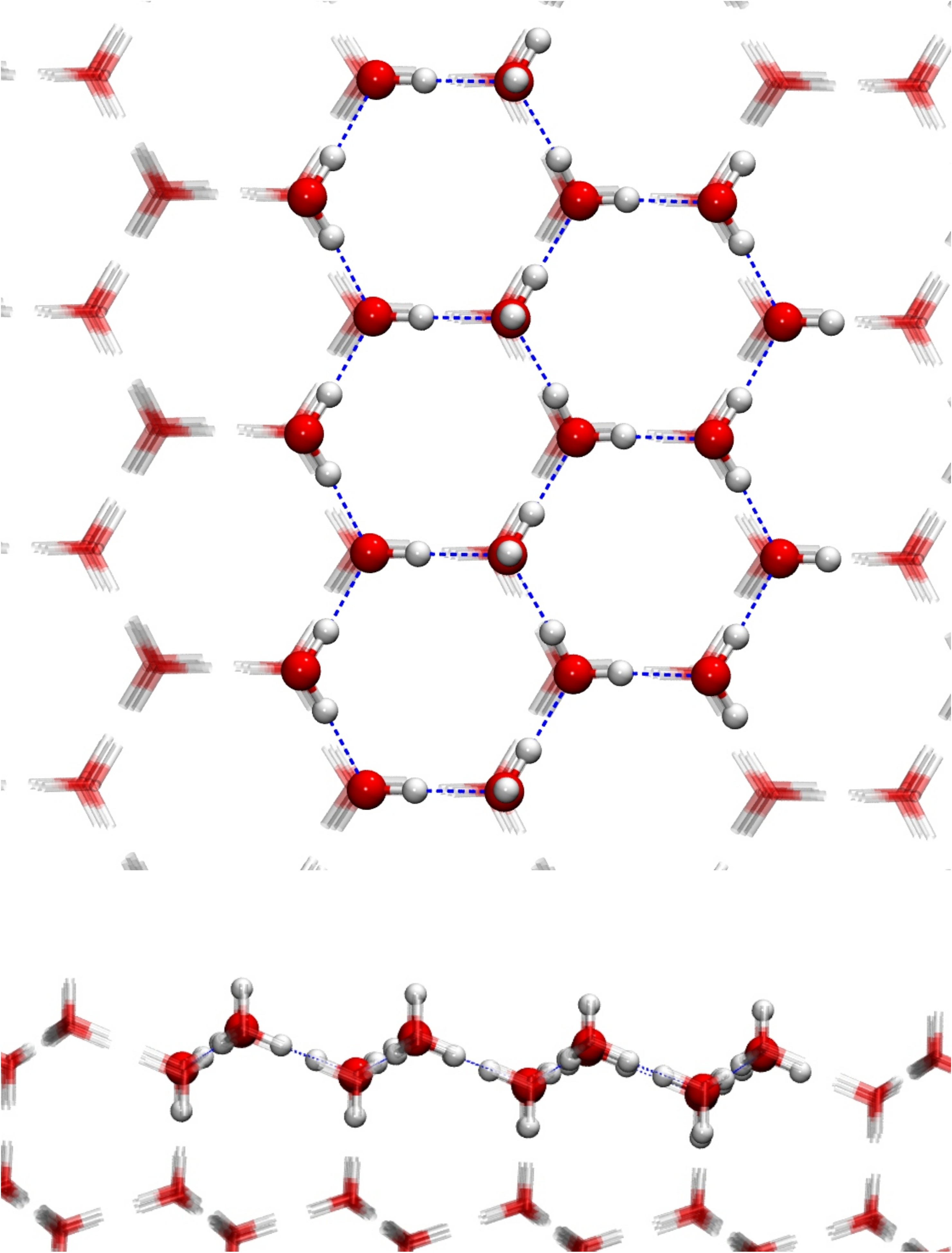}
  \caption{QM/MM setup of the Fletcher surface.
 The solid ball-and-stick-molecules represent the QM region while the transparent ones represent the MM region. All MM water molecules within a radius of 15\,\AA\, are allowed to move and rotate.
    \label{fig:h3o:QMMM}
  }
\end{center}
\end{figure}

We used the (0001) surface of hexagonal ice I$_\text{h}$ that
minimizes the surface free energy 
as described by Fletcher.\cite{Fletcher1992} 
The structure of the bare surface can be seen in \figref{fig:h3o:QMMM}.
In that phase, the
protons are ordered. Equivalent sites for rows (vertical in the top
image in \figref{fig:h3o:QMMM}). While the majority of solid water in
the ISM is expected to be amorphous, the crystalline phase is easier
to model because only a limited number of distinct adsorption sites is
available. 

The structural model consisted of a hemisphere with a radius of 
{ 25 \AA{} comprising 1151 water molecules}. 
The MM part was described by the TIP3P potential.\cite{jor83}
For the QM part, we used five adjoining water hexagons of the top layer:
19 water molecules and the adsorbed atoms.
Geometry optimisations included
all water molecules within a radius of 15\,\AA \, around the center of
the hemishphere (261 molecules) as active atoms, the other H$_2$O molecules were frozen. 
This leads to a proton disordering of the surface.
All in all, 2349 variables are to be optimised for the bare surface.
Instantons and the corresponding Hessians have been calculated with a reduced dimensionality:
here, only the adsorbed atoms/molecules and the hexagon of six closest
water molecules were 
flexible.

\subsection{Reaction Rate Calculations and Tunneling}

To calculate reaction rate constants including atom tunneling
we use instanton theory, 
\cite{aff81,col88,han90, ben94,mes95,ric09,alt11,rom11,rom11b,ric16} 
a method based on Feynman's path integrals,\cite{fey48}
which is increasingly used\cite{kae14}
to calculate chemical reaction rate constants.\cite{cha75,mil94,mil95,mil97,sie99, sme03,qia07,and09,
  gou10a,gou11,gou11b, rom11,gou10,jon10,mei11,gou11a,ein11,rom12,
  kry12,kae13,alv14,kry14,lam16,son16,alvarez2016,lamberts2017,kob17} 
In instanton theory, a closed Feynman path spans the barrier region.
At low temperatures it extends towards the reactant state.
At temperatures above the crossover temperature
\begin{equation}
 T_\text{c}  = \frac{\hbar \omega_\text{TS}}{2 \pi k_\text{B}}
 \label{eq:h3o:2}
\end{equation}
the instanton path generally collapses to one single point on the
 potential energy surface.\cite{alv14}
Here, $\hbar$ is Planck's constant devided by $2\pi$, $k_{\text{B}}$
is Boltzmann's constant and $\omega_{\text{TS}}$ is the 
absolute value of the imaginary frequency at the transition state structure 
in the harmonic approximation.
The crossover temperature gives a first and simple estimate
at which temperature atom tunneling becomes important.
The mass-dependence of $\omega_{\text{TS}}$  
also causes mass-dependence of $T_\text{c}$.

For the reaction on the Fletcher surface, the closed Feynman path was discretized 
with  40 images down to 175\,K and 
with  78 images down to  80\,K.
For the gas phase reaction (of all isotopologues),
200 images were used for the whole temperature range.
Vibrational modes are included harmonically around the Feynman path.
The rotational partition functions of the reactants 
and the images of the instantons were approximated by those of rigid rotors.
The translational partition function was included within the approximmation of an ideal gas,
which is 
identical to the quantum particle in a box.
The rotational partition function of the whole instanton was calculated to be the 
geometric mean value of the rotational partition functions of all images.
The symmetry number $\sigma=2$ was taken into account when calculating the 
rotational partition function of the H$_2$ and D$_2$ molecules for bimolecular reaction rates.\cite{fer07a}
For the reactions with HD and for unimolecular reaction rates in general, $\sigma = 1 $ was used
because the rotation of adsorbed molecular hydrogen is hindered by the
surface.

Besides the structural model which includes the surface atoms
explicitly, we alternatively mimic the effect of the  ice surfaces on the partition function
for reactions calculated in a gas-phase model in an 
approach we sucessfully applied previously.\cite{lamberts2017}
For unimolecular reaction rate constants, the rotational partition function 
is assumed to be constant during the reaction just as the surface
surpresses the rotation in the reactant as well as in the transition state.
For bimolecular reaction rate constants, only the translation and rotation of the H$_2$ fragment 
is considered in the reactant state.
Rotational and translational motion of the OH radical and transition
state structure are suppressed, just it is the case when OH is
adsorbed on the surface.
It has to be mentioned that while this approach properly approximates
the suppressed motion of the species on the surface, it neglects any
influence of the surface on the potential energy along reaction path
and therefore the corresponding potential activation energy.
Hereinafter, this methodology is referred to as \emph{implicit surface model}.

Instantons were considered to be converged when all components of the nuclear gradient 
are smaller than $1 \cdot 10^{-8}$ a.u..
Instanton calculations were performed at temperatures below the crossover temperature
of 280\,K $\pm$ 5 K (depending on the binding site).
Because of the existance of a pre-reactive minimum,
below a particular, mass-dependent temperature, the tunneling energy
is lower than the potential energy of the separated products for
bimolecular reaction rates. At that temperature, canonical instanton
theory becomes unreliable.\cite{mcconnell2017} 
Therefore, bimolecular reaction rates can only be provided for 110\,K and higher temperatures for the reaction H$_2$ + OH $\rightarrow$ H + H$_2$O.

Calculations of intrinsic reaction coordinates (IRCs) have been performed
using a modified version of Schlegel's Hessian-Predictor-Corrector
method\cite{hra04,hra05,meisner2017inprep} with a step size of 0.05
mass-weighted atomic units.

\section{Results}
\subsection{Binding Sites and Energies}

  \begin{figure*}[bt]
  \begin{center}
\includegraphics[width=15cm]{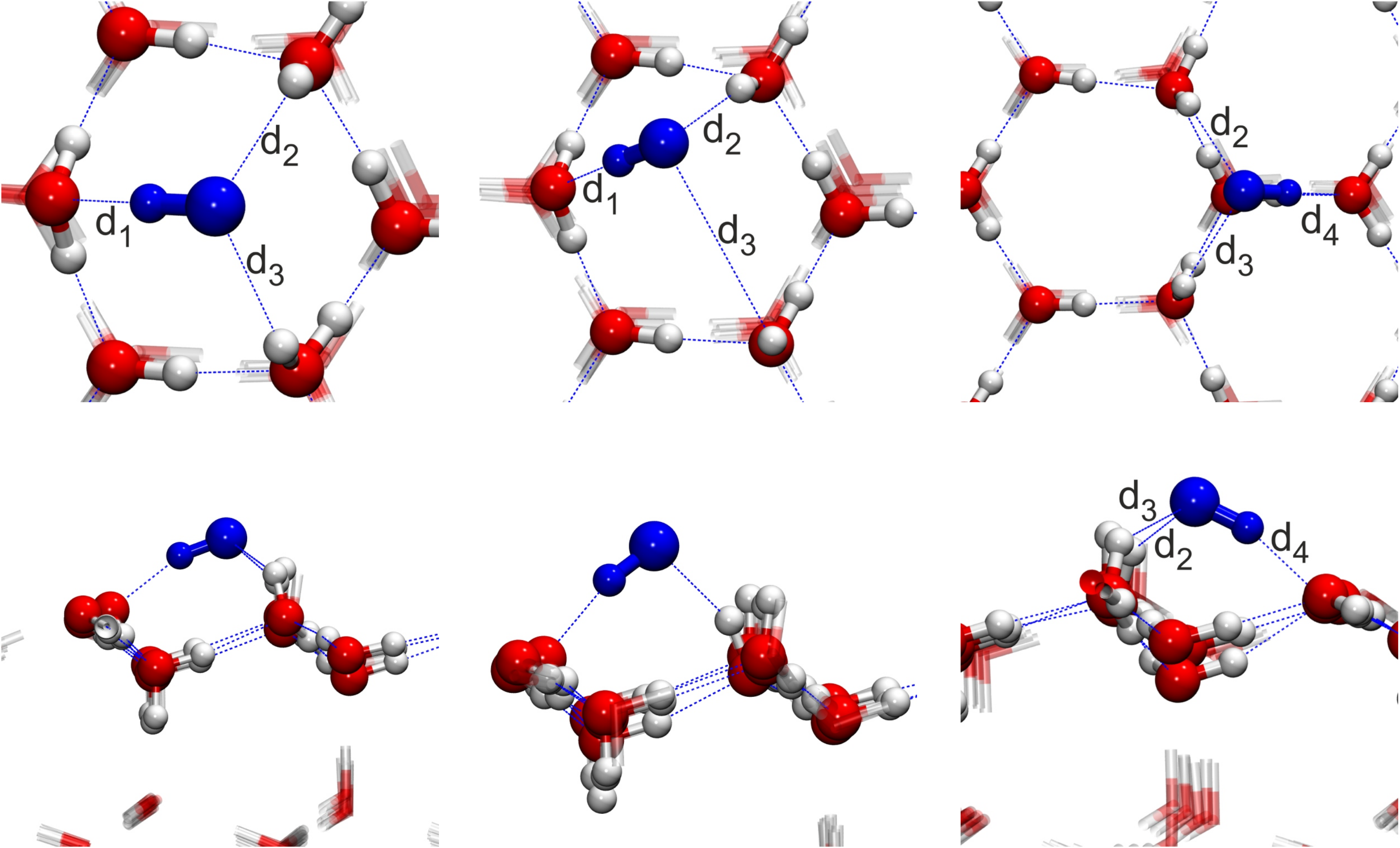}
    \caption{
      Structures of the adsorbed OH radical (blue) on the Fletcher surface.
      All binding sited are shown from top and side perspectives.
      Left:    \emph{Hollow},
      Middle: \emph{Bridged},
      Right: \emph{Top} 
      \label{fig:h3o:OH_adsorbed}
    }
  \end{center}
  \end{figure*}
We identified three different binding sites of OH on the Fletcher surface,
which are shown in \figref{fig:h3o:OH_adsorbed}.
We calculated the corresponding adsorption energies with and without harmonically approximated
vibrational zero-point energies.
The values are given in \tabref{tab:h3o:oh:adsorbtion}.
For all binding sites, zero-point energy reduces the binding energy by 
around 16\,kJ\,mol$^{-1}$ (1920\,K) because the OH--surface complex has additional vibrational modes.
The harmonic approximation can be assumed to overestimate the zero-point energy 
which leads to an underestimation of the corresponding adsorbtion energy.

In the first binding site, the OH radical is located directly in the middle of a 
water hexamer where it accepts hydrogen bonds from two of the dangling hydrogen atoms and donates a hydrogen bond to the O atom of a water molecule of the surface.
We call this binding site \emph{hollow}.
The binding energy is  40.5\,kJ\,mol$^{-1}$ (4870\,K) including zero-point energy.

In a similar binding site the OH radical is also hydrogen-bound to 
the oxygen atom of a water molecule 
and to one dangling hydrogen atom of the surface, see \figref{fig:h3o:OH_adsorbed}.
The third hydrogen bond is absent, \emph{i.e.,} the OH radical 
bridges two surface water molecules.
We call this binding site \emph{bridged}. 
As this binding site is rather similar to the \emph{hollow} one, the 
binding energy is with 39.7\,kJ mol$^{-1}$ (4770\,K) only slightly smaller.

In the third binding site the OH radical is located on top of one of the water molecules. 
Because of that we call this binding site \emph{top}.
The OH radical also accepts two hydrogen bonds from the surface and donates one.
The binding energy of this site is with 32.1\,kJ mol$^{-1}$ (3860\,K) lower than for the \emph{hollow}
and \emph{bridged} binding sites.
These values lie nicely in the range of experimentally determined 
desorption energies on silicate surfaces of
1656--4760\,K.\cite{he2014}

\begin{table}[h!]
 \caption{
Adsorption energies of OH on the Fletcher surface and hydrogen bond distances.
$V_{\text{ads}}$ and $E_{\text{ads}}$ denote the adsorbtion energy without and with 
zero point energy, respectively.
The hydrogen-bond legths $d_1$ to $d_4$ are explained in \figref{fig:h3o:OH_adsorbed}.
Energies are given in kJ mol$^{-1}$, distances in \AA.
 \label{tab:h3o:oh:adsorbtion}}
    \begin{center}
\begin{tabularx}{0.96\columnwidth}{XXXXXXX}
\hline
&$V_{\text{ads}}$ & $E_{\text{ads}}$
&  $d_1$
&  $d_2$
&  $d_3$
&  $d_4$ \\
\hline
Hollow 			&57.0	&40.5	&	1.77	& 2.22	& 2.27	&	\\
Bridged			&55.6	&39.7	& 	1.77	& 1.99  & 3.89	&	\\
Top			&48.1	&32.1	&      		& 2.11	& 2.36	& 1.78	\\ 
\hline
\end{tabularx}
\end{center}
\end{table}

\subsection{Reaction Barriers}
One possibility for the 
reaction of the OH radical with molecular hydrogen on the ice surface is 
the Eley--Rideal (ER) mechanism,
in which one species is adsorbed on the surface and the other one is
approaching from the gas phase.

We have shown that the binding energy of OH radicals is 
much higher than the
binding energy of H$_2$ molecules (3.6--4.6\,kJ mol$^{-1}$; 440--555\,K).\cite{cuppen2017}
Therefore, we first investigate the reaction of adsorbed OH radicals 
with H$_2$ molecules directly from the gas phase:
\begin{reaction}
\text{OH}_{\text{ads}} + \text{H}_{2(g)} \rightarrow \text{H}_2\text{O} + \text{H}.
\label{rxn:h3o:ER}
\end{reaction}
In this work, we restrict ourselves to the chemical reaction forming H$_2$O
molecules and ignore the physical processes after that. 
Therefore, for the products the label indicating the aggregate state was omitted. 
For each binding site of OH we found one corresponding transition state structure.
These transition state structures are called 
\emph{direct} hereinafter.
The vibrational adiabatic reaction barriers 
with respect to the separated reactants  
(potential energy barriers including zero-point energy)
of these transition state structures
lie between 24.2 and 24.7\,kJ mol$^{-1}$, (2910 and 2970\,K), see \tabref{tab:h3o:oh_barriers}.
This is just slightly lower than the 
adiabatic reaction barrier of the gas phase reaction of 
25.4\,kJ mol$^{-1}$ (3055\,K, from separated reactants).
Here we want to stress that all reaction barriers are very similar, independently of the corresponding adsorbtion energies, although the latter 
vary over 8\,kJ mol$^{-1}$. This is in agreement with what is found for
reactions on amorphous solid water.\cite{son16}

\begin{figure}[bth]
  \includegraphics[width=8cm]{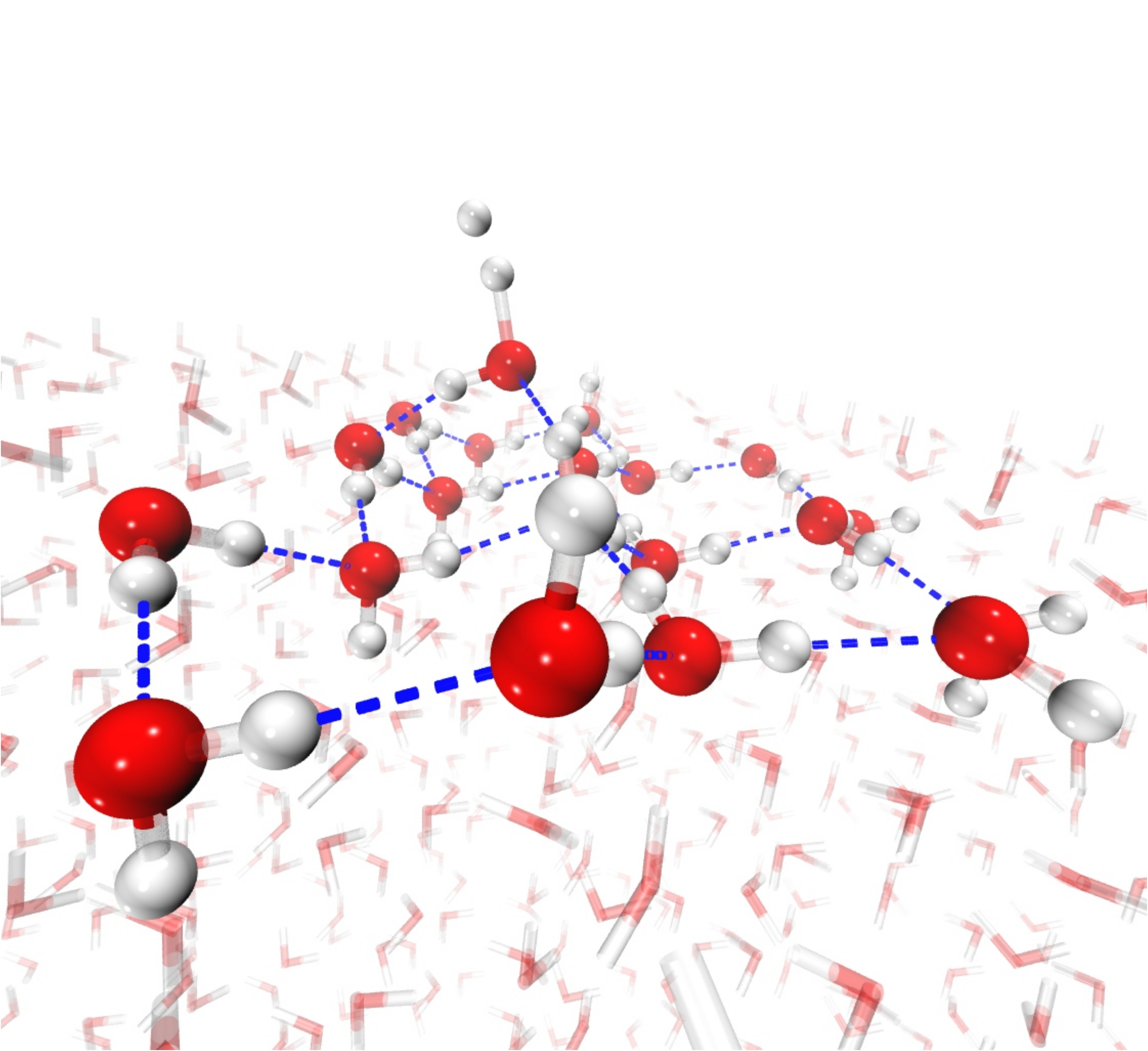}
   \caption{Transition state structure of the \emph{direct-hollow} binding
     site. QM atoms are shown as balls and sticks, MM atoms as transparent
     sticks. 
     \label{fig:h3o:TS}
   }
\end{figure}

We found another type of transition state structures 
in which the hydrogen atom of the OH radical points away from the surface
and the H$_2$ molecule approaches via a path closer to the surface.
The [OH $\cdots$ H$_2$] complex possesses similar internal
coordinates as in the \emph{direct} transition state structures
but is rotated with respect to the
ice surface.
For these \emph{rotated} transition state structures, 
the adiabatic reaction barriers 
are 49.3 and 45.7\,kJ mol$^{-1}$
(5930\,K and 5500\,K) 
for the \emph{hollow} and the \emph{top} binding site, respectively.
No \emph{rotated}
transition state structure was found for the \emph{bridged} binding
site. Since the barriers via \emph{rotated} transition states are much higher
than those via the \emph{direct} transition states, the latter have result in
much higher rates and the former are not considered further.

\begin{table}[bt]
 \caption{
Reaction barriers including zero-point energies.
The label \emph{bi} 
denotes that the barrier is given 
with respect to the separated reactants, \emph{i.e.},
OH$_{(ads)}$ and  H$_{2(g)}$.
The label \emph{uni} indicates 
barriers with respect to the respective pre-reactive complexes.
All values are in kJ mol$^{-1}$.
 \label{tab:h3o:oh_barriers}}
    \begin{center}
\begin{tabularx}{0.96\columnwidth}{XXXXX} 
\hline
&{Hollow} 
&{Bridged   }
&{Top    } 
&Gas Phase \\
\hline
$E_{A,\text{bi}}^{\text{direct}}$		&24.2	&24.7	&24.3	&25.4	\\
$E_{A,\text{bi}}^{\text{rotated}}$		&49.3	&---	&45.7	&---	\\
$E_{A,\text{uni}}^{\text{direct}}$ 		&22.5	&24.1	&22.4 	&24.4	\\
\hline
\end{tabularx}
\end{center}
\end{table}

\begin{table}[bt]
 \caption{
Geometric parameters of the
transition state structures for different binding sites in comparison to 
the gas phase transition state structure.
The O--H distance of the newly formed bond, 
the H--H distance of the original H$_2$ molecule, 
and the H--O--H angle of the newly formed water molecule are denoted by
d$_{\text{O--H}}$,
d$_{\text{H--H}}$, and  
$\measuredangle$(H--O--H), respectively. 
Distances are in \AA{} and angles in degrees.
 \label{tab:h3o:}}
    \begin{center}
\begin{tabularx}{0.95\columnwidth}{XXXX} 
\hline
&
d$_{\text{O--H}}$		&
d$_{\text{H--H}}$		&
$\measuredangle$(H--O--H)		\\
\hline
Hollow 			&1.33	&0.83	&99.2		\\
Top			&1.33	&0.83	&99.7		\\ 
Bridged			&1.33	&0.84	&98.7		\\
Gas phase 		&1.36	&0.82	&96.8		\\
\hline
\end{tabularx}
\end{center}
\end{table}

We calculated the potential energy along the intrinsic reaction coordinates (IRCs) 
of the different binding sites. 
The end of the IRCs define pre-reactive complexes (PRCs). 
Those are geometries in which an H$_2$ molecule 
is loosely bound to the surface in the vicinity of the OH radical.
These structures are used as reactant states 
to calculate the unimolecular activation energies shown in \tabref{tab:h3o:oh_barriers}.

\begin{figure}[!h]
       \includegraphics[width=8cm]{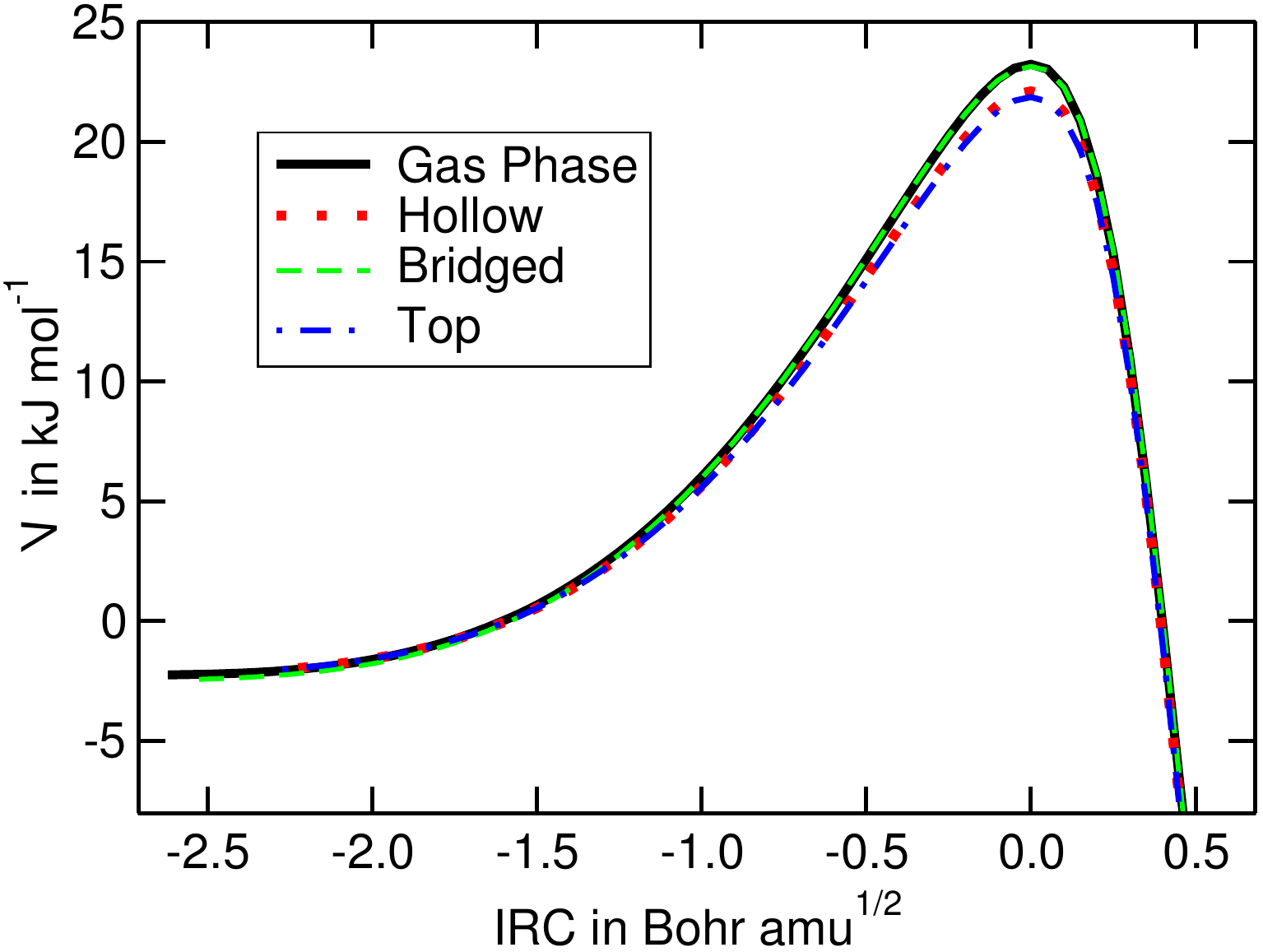}
\caption{
    Above: 
    Potential energy $V$ of the intrinsic reaction coordinates (IRCs)
    of the \emph{hollow}, \emph{top}, and the \emph{bridged} binding site 
    compared with the gas-phase reaction.
\label{fig:h3o:ircbenchmark}
}
\end{figure}

The potential energy curves along the IRCs belonging to the different binding sites 
are almost indistinguishable from the one of the gas-phase reaction, see \figref{fig:h3o:ircbenchmark}.
This shows again that the surface has neglibible influence on the potential energy along the 
reaction path.
During the reaction, any changes in the hydrogen bond length remain below 0.2
\AA, see Fig. S3 of the Supplementary Information.
Overall, a classical catalytic effect is absent: the activation barrier is unaltered, neither is the reaction mechanism changed.  
This can be explained by the adsorption energy during the reaction.
In typical heterogeneous catalysis
the molecules are activated through interactions with the surface in a way
that the energy of the transition state with respect to the energy 
of the reactant state is reduced.
These interactions are either forming new chemical bonds or causing a shift in electron density.
In our case, the OH radical forms three hydrogen bonds in the adsorption
process and these three hydrogen bonds are retained during the whole reaction.
Therefore, the adsorption energies of the
reactant state and the transition state structure 
are virtually the same and the potential energy of the reaction remains comparable to the
gas phase. 
Reaction with an OH radical bound via four hydrogen bonds is impossible due to
steric hinderence.
Therefore, the maximal number of H-bonds to a reactive OH is always three, independent of the existence of 
\emph{e.g.} cavities for surface defects.
It can, therefore, be assumed that amorphous solid water ices 
behave similarly in terms of negligible catalytic effect.

Note, that any processes after the formation of the chemical bonds, like
desorption or dissipation, are outside of the scope of this article because
the do not influence the rate.
The kinetic bottleneck in the water formation from H$_2$ molecules 
and OH radicals is the H--H bond breaking which is described here.

\subsection{Reaction Rate Constants for the Eley--Rideal Mechanism}
In the Eley--Rideal (ER) mechanism, one particle (a molecule or an atom) physisorbs on the surface
and thermalizes there. Another particle comes and
directly reacts with the pre-adsorbed particle to form the products.
In this study, we want to focus on  \rxnref{rxn:h3o:ER}
where an OH radical is adsorbed 
and the H$_2$ molecule comes in from the gas phase, since 
OH has a higher adsorption energy.
The incoming H$_2$ molecule reacts
with the OH-surface system 
in what can be formally seen as a bimolecular reaction.
Instantons were calculated from 250\,K to 110 K.
For comparison, we calculated reaction rate constants using the approximation 
of an Eckart-shaped barrier.

As the reaction profiles of all three 
\emph{direct} transition state structures
and IRCs are nearly identical, we 
only calculated reaction rate constants of the 
\emph{direct-hollow} transition state structure.
Due to high computational costs the active region for the instanton calculations 
was reduced to the one water hexamer below the adsorbed OH radical.
The resulting adiabatic activation energy of 24.11\,kJ~mol$^{-1}$ almost equal
to the 24.19\,kJ~mol$^{-1}$
 obtained for the full active region.  

\begin{figure}[tb]
\begin{center}
  \includegraphics[width=8cm]{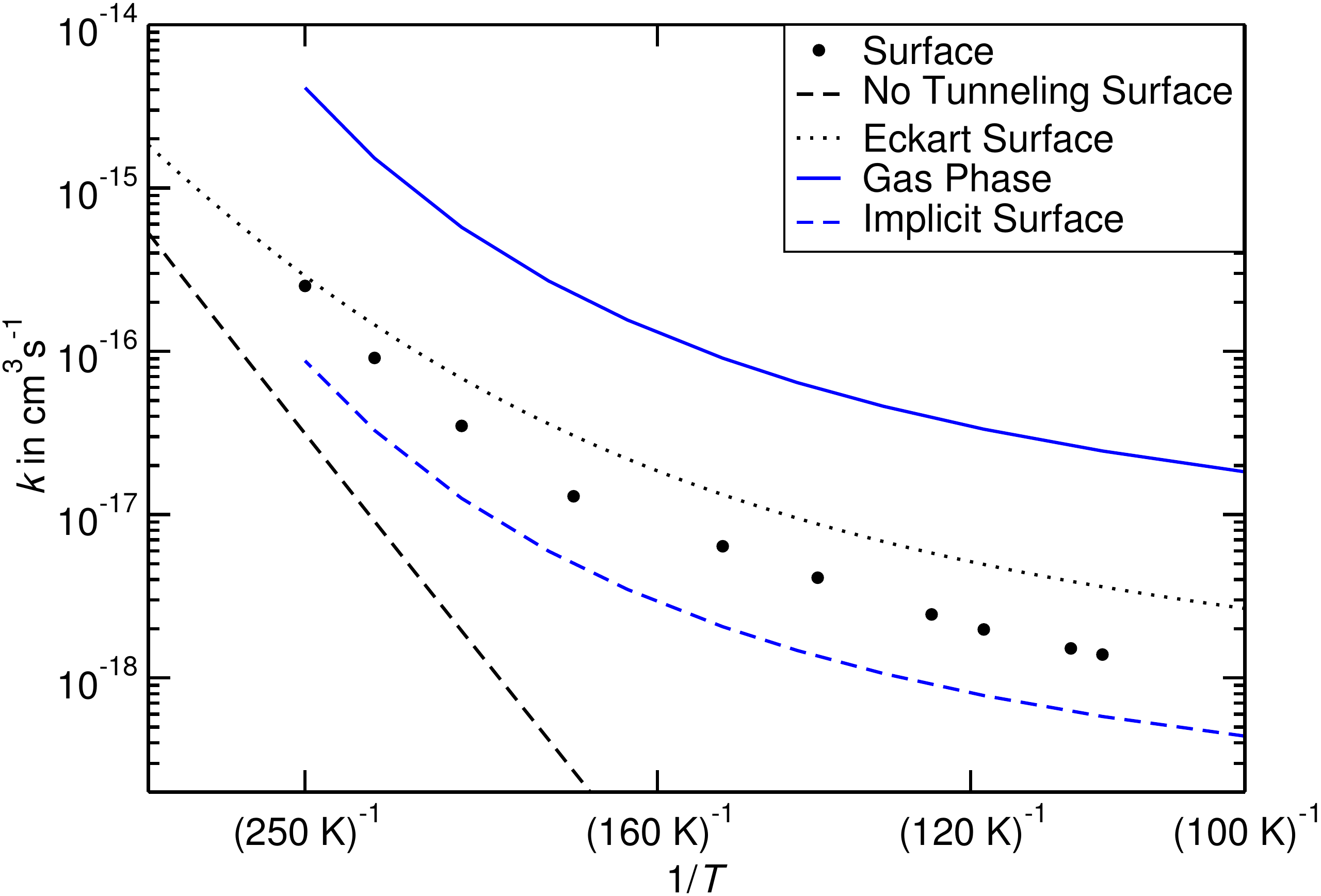}
  \caption{Arrhenius plot of the bimolecular (Eley--Rideal) reaction rate constants. 
           Instanton theory is used if not stated otherwise. 
           Surface reaction rates are calculated for the \emph{direct-hollow} reaction path.
  \label{fig:h3o:rates_bimol}
  }
\end{center}
\end{figure}

The resulting reaction rate constants are
compared to the gas-phase data
calculated on the same potential energy surface
in the Arrhenius plot
in \figref{fig:h3o:rates_bimol}.
The rate constants including tunneling correction via the 
Eckart barrier and those obtained by transition state theory without tunneling are shown for comparison.
The implicit surface model is able to reproduce the rate constants of the
explicit surface calculations up to factors of 3.6 and 2.7 at 275\,K and  and 110\,K, respectively.
This indicates that for reactions without a classical catalytic effect,
the implicit surface approach is a promising approximation. Numeric values for
the rate constants are available in the Supplementary Information, table S II.

\subsection{Reaction Rate Constants for the Langmuir--Hinshelwood Mechanism}
In the Langmuir--Hinshelwood (LH) mechanism, 
both particles are adsorbed on the surface and diffuse until they meet. If
they approach each other, they form a PRC. This PRC can either react or decay
by diffusion or desorption of one or both reactants. The reaction of a PRC to
the products is a unimolecular process. Thus, Langmuir--Hinshelwood reactions
are characterized by unimolecular rate constants.

\begin{figure}[t!b]
\begin{center}
  \includegraphics[width=8cm]{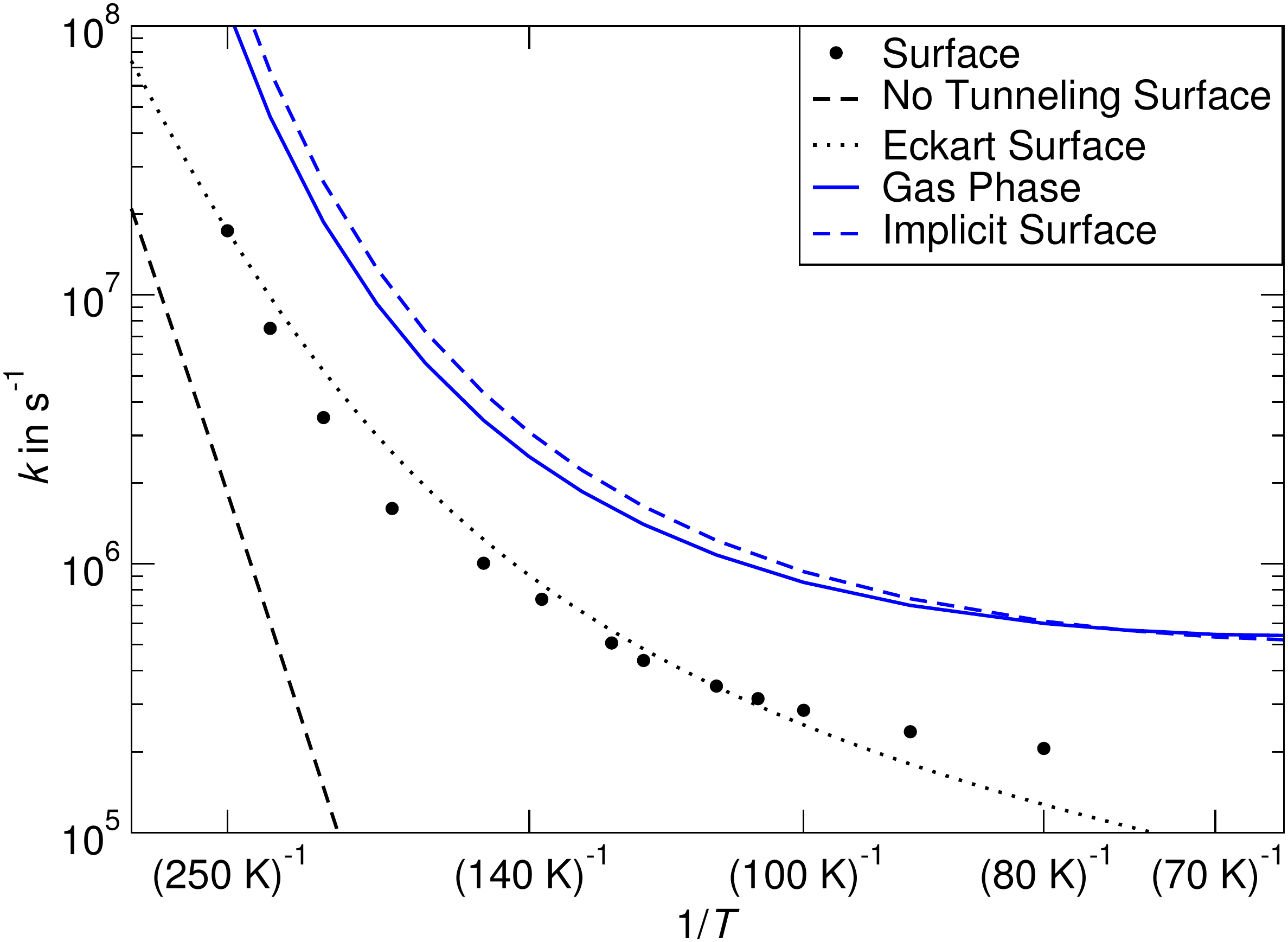}
  \caption{Arrhenius plot of the unimolecular Reaction rate constants.
           Surface reaction rates were calculated around the direct-hollow reaction pathway.
  \label{fig:h3o:rates_unimol}
  }
\end{center}
\end{figure}

We calculated unimolecular reaction rate constants 
for the \emph{hollow} binding site.
The adiabatic activation barriers for the LH mechanism 
in all binding sites are given as $E_{A,\text{uni}}^{\text{direct}}$ in \tabref{tab:h3o:oh_barriers}.
The resulting unimolecular rate constants $k_{\text{react}}$ are shown in 
\figref{fig:h3o:rates_unimol} and table S~II.
Instantons were calculated down to 80\,K. At even lower temperatures, more 
images would be required to obtain converged reaction rates which 
would render the computations too expensive.

The rate constants from the implicit surface model 
agree within one order of magnitude with those from the full ice surface model,
see \figref{fig:h3o:rates_unimol}.

The main effects of a surface on catalysis are 
\begin{enumerate}
\item  An increase in the concentration of reactive species compared to the gas phase, especially in low-pressure environments like the ISM.
\item  The removal of excess heat of reaction and, thus, the stabilization of reaction products of exothermic reactions. 
\item  Restricted mobility, in particular rotation.
\item  Alternations of the barrier height and possibly the reaction path, \emph{i.e.,} a classical catalytic effect. 
\end{enumerate}
In any atomistic models based on transition state theory,
as used in the present work, effects (1) and (2) are included implicitly. Rate
constants are independent of the concentrations. Thermal rate constants assume
a canonic ensemble, \emph{i.e.,} thermal equilibrium throughout the reaction. Excess
heat is removed instantly.
The implicit surface model we propose here also
includes (3), the immobilization. Only (4), the classical catalytic effect, is
neglected by the implicit model, but taken into account in an explicit surface
model in which the surface atoms are actually included in the structural model.

The results of the standard gas-phase model and the gas-phase calculations
using the implicit surface model are closer for the unimolecular reaction than
for the bimolecular calculations. 
The reason is that in the unimolecular case, the implicit surface model merely
assumes that the rotational partition function of the PRC is the same as the
one of the transition state (\emph{e.g.} that both do not rotate), while their
rotation is taken into account in the gas phase. As both PRC and transition
state have similar rotational partition functions
(they include the same atoms), the neglect of this term is of minor effect.
In the bimolecular
case, the implicit surface model removes translation and rotation of one
reactant (OH) and the transition state, which is a much larger alternation of
the rate constant.
Note, that due to the inhibited rotation of the H$_2$ molecule in the PRC 
the symmetry number $\sigma=1$ was used for the explicit and implicit surface calculations.

\subsection{Kinetic Isotope Effects}
\begin{figure}[bt]
\begin{center}
  \includegraphics[width=8cm]{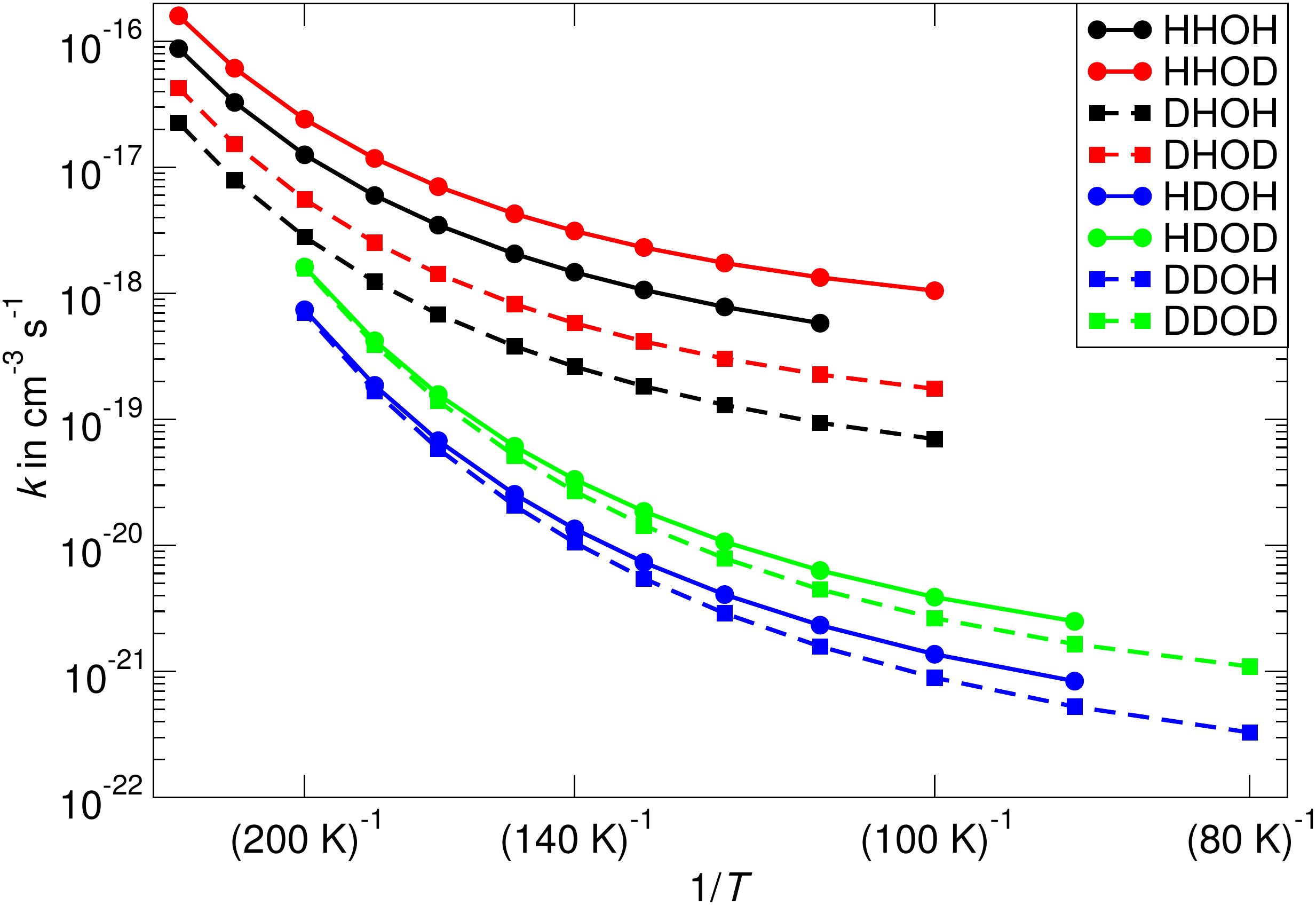}
  \caption{
     Temperature dependence of the (bimolecular) Eley-Rideal reaction rate constants of all H/D isotopologues calculated with the instanton method and the implicit surface approximation.
    \label{fig:h3o:bimol_isotopes_norotation}
  }
\end{center}
\end{figure}

\begin{figure}[t!]
\begin{center}
  \includegraphics[width=8cm]{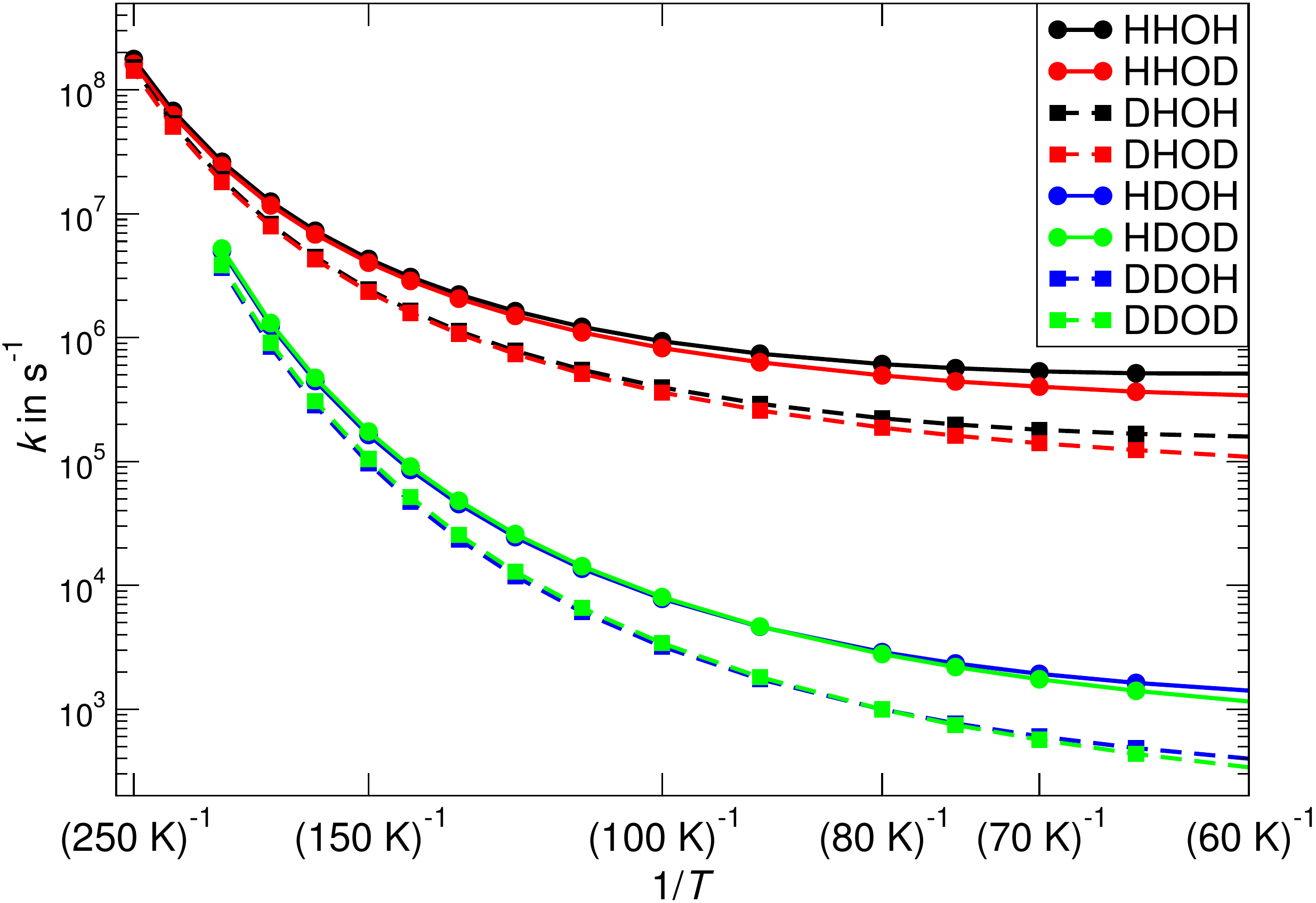}
  \caption{
     Temperature dependence of the unimolecular 
     rate constants of all H/D isotopologues.
     which can be used to calculate $R_{\text{LH}}$ 
     calculated with the instanton method and the implicit surface approximation.
    \label{fig:h3o:unimol_isotopes_norotation}
  }
\end{center}
\end{figure}

We used the implicit surface model to calculate 
kinetic isotope effects (KIEs) for all eight possible deuteration patterns.
Bimolecular and unimolecular reaction rate constants are shown in 
\figref{fig:h3o:bimol_isotopes_norotation}
and
\figref{fig:h3o:unimol_isotopes_norotation},
respectively. 
When substituting protium atoms with deuterium atoms, the crossover
temperature reduces significantly as a result of a smaller imaginary frequency and therefore a smaller crossover temperature. Thus, the rate constants are reported here for 200\,K and below.
For bimolecular reaction rates, 
the temperature below which the tunneling energy is lower than the potential
energy of the asymptotic reactants changes with the mass, too.
Here and in
\figref{fig:h3o:bimol_isotopes_norotation} and
\figref{fig:h3o:unimol_isotopes_norotation} the isotope patters are labeled as
in our previous work\cite{mei16a} as
H$^1$H$^2$OH$^3$ such that the reaction reads
$\textnormal{H}^1\textnormal{H}^2+ \textnormal{O}\textnormal{H}^3 \rightarrow
\textnormal{H}^1 + \textnormal{H}^2\textnormal{O}\textnormal{H}^3$. DDOH
therefore corresponds to a reaction of OH with D$_2$ while HDOH corresponds to
the reaction HD + OH $\rightarrow $ H + DOH.

In both, the bimolecular and unimolecular cases, the primary KIE 
-- the ratio between the rate constants of H-transfer and the corresponding D-transfer --
is as big as two orders of magnitude but also depends on the isotope pattern of the other two hydrogen atoms.

For the bimolecular case, the KIEs are similar to the ones reported in the gas phase.\cite{mei16a}
The secondary KIEs play a smaller role as they are in all cases smaller than 10.
When substituting OH by OD, an inverse KIE is found, \emph{i.e.,} the rate constant
increases due to deuteration. 
This small inverse secondary KIE, about
2--3 is caused by differences in the zero-point energy. 
\cite{per12a,sulemanov2013, per14,mei16} 
The inverse KIE for the deuteration of OH was also present in the gas-phase reaction rate constants 
where the difference in zero-point energy corrected activation barrier 
between the HHOH and the HHOD system is
1.3\,kJ mol$^{-1}$ (156\,K).\cite{mei16a}

In the unimolecular case, the primary KIEs are $\approx 5$ at 200\,K and increase to  $\approx 300$ at 60\,K.
The secondary KIEs play an even smaller role than in the bimolecular case 
and no inverse KIE is present, see \figref{fig:h3o:unimol_isotopes_norotation}.

\section{Astrochemical Implications}
In order to provide our calculated rate constants to astrochemical modelers in an easily implementable way, we have fitted all curves in Figs.~\ref{fig:h3o:bimol_isotopes_norotation} and~\ref{fig:h3o:unimol_isotopes_norotation} to the following analytical expression:\cite{zhe10}
\begin{equation}
k(T) = \alpha   \left(\frac{T}{300 \text{ K}}\right)^{\beta}\exp\left(-\gamma\frac{T+T_0}{T^2+T^2_0}\right)
\label{eq:h3o:zheng}
\end{equation}
Here, the parameters $\alpha$, $\beta$, $\gamma$, and $T_0$ are all fitting parameters, where $\alpha$ has the units of the rate constant, $\beta$ regulates the low-temperature behavior, and $\gamma$ and $T_0$ can be related to the activation energy of the reaction. To obtain a realistic low-temperature extrapolation the value of $\beta$ has been fixed to 1.
For the fitting procedure we use the Eckart approximation above the crossover temperature, \eqref{eq:h3o:2}, and the values calculated with instanton theory below $T_\text{c}$. 
The exact values for the fit parameters for all eight isotope-substituted reactions are given in the Supplementary Information along with the corresponding reaction rate constants. 

We have found that the energy barrier is rather independent of the binding site. We have found the same trend for the reaction of HNCO with hydrogen atoms on amorphous solid water.\cite{son16} Furthermore, from the figures in the previous Section it is apparent that the curves of the rate constants flatten off with decreasing temperature. Therefore, in order to be able to use our unimolecular rate constants in models that take the very low temperatures in dense molecular clouds (20~K) into account, we recommend to use the value of the rate constant at the lowest temperature given (60~K) with an approximate error bar of $\pm$ half an order of magnitude. 

As mentioned above, there is a relation between bimolecular rate constants and the Eley--Rideal mechanism and unimolecular rate constants and the Langmuir--Hinshelwood mechanism. Given the typical low fractions of OH radicals available at the surface for direct H$_2$ impingement,\cite{lam14b,Chang:2014} the Langmuir--Hinshelwood mechanism where two species find each other via surface diffusion is expected to dominate. 


 \begin{table*}
 \caption{Reaction rate constants $k$ in cm$^3$ s$^{-1}$ and kinetic isotope effects for low-$T$ unimolecular reactions with
   different approaches. 
 \label{tab:h3o:kie}}
\begin{tabular}{l@{\ \ \ }cc@{\ \ \ \ \ }cc@{\ \ \ }c}
        \hline
          & \multicolumn{2}{c}{Rectangular barrier} & \multicolumn{2}{c}{Eckart barrier} & Instanton theory\\
        Rate constant  &
        ref\citenum{fur13}$^{\dagger}$ &            this work$^{\ddagger}$ &             ref\citenum{taq13}$^a$&
                  this work$^{\circ}$ &                 this work$^{\circ}$\\
        \hline
        HHOH                         &  $1.40\times10^{+1}$    & $6.31\times10^{+2}$   & $4.07\times10^{+5}$   & $3.19\times10^{+5}$    &    $5.12\times10^{+5}$   \\
        DHOH                         &  $1.11\times10^{-1}$    & $6.31\times10^{+2}$   & $3.62\times10^{+5}$   & $1.16\times10^{+5}$    &    $1.59\times10^{+5}$   \\
        HDOH                         &  $1.11\times10^{-1}$    & $9.76\times10^{-2}$   & $1.00\times10^{+3}$   & $3.12\times10^{+2}$    &    $1.41\times10^{+3}$   \\
        HHOD                         &  $1.30\times10^{+1}$    & $6.31\times10^{+2}$   & $8.74\times10^{+5}$   & $4.62\times10^{+5}$    &    $3.42\times10^{+5}$   \\
        DDOH                         &  $2.51\times10^{-3}$    & $9.76\times10^{-2}$   & $8.07\times10^{+2}$   & $3.40\times10^{+2}$    &    $3.99\times10^{+2}$   \\
        \hline
        KIEs \emph{wrt.} HHOH \\
        \hline
        DHOH       &  127   & 1.00  & 1.12  &  1.37        &       3.22\\
        HDOH       &  127   & 6465  & 407   &  511 &       363\\
        HHOD       &  1.08  & 1.00  & 0.466 &  0.69        &       1.50 \\
        DDOH       &  5578  & 6265  & 504   &  938 &       1283\\
        \hline
 \end{tabular}
\begin{flushleft}
$^{\dagger}$ $a=1$~\AA, $m$ as the reduced mass, $E_\text{reaction}=2100$~K\\
$^{\ddagger}$ $a=1$~\AA, $m$ as the mass of the transferring atom, $E_\text{reaction}=2700$~K \\
$^{\circ}$ at 60~K\\
$^a$ Literature values \cite{taq13} multiplied by $\nu_\text{trial}=10^{12}$
  s$^{-1}$ 
\end{flushleft}
 \end{table*}

The Langmuir--Hinshelwood mechanism can be described as a reaction cascade:
\[
\text{OH}_{\text{(ads)}} + \text{H}_{2\text{(g)}}  
\rightarrow \text{OH}_{\text{(ads)}} + \text{H}_{2\text{(ads)}}  
\]
\begin{reaction}
\xlongrightarrow[\text{}]{k_{\text{diff}}}
[\text{OH}\cdots \text{H}_2]_{\text{(ads)}}
\xlongrightarrow[\text{}]{k_{\text{react}}}
\text{H}_2\text{O} + \text{H}
\label{rxn:h3o:LH}
\end{reaction}
Diffusion forms the pre-reactive complex (PRC), which reacts to the products. The reaction rate of the last step is $R_\text{react}=k_{\text{react}} n([\text{OH}\cdots \text{H}_2]_{\text{(ads)}})$, 
\emph{i.e.,} a unimolecular process with the rate constant $k_{\text{react}}$. The rate of the overal LH process, $R_\text{LH}$, can be expressed as the probability to react, $P_\text{react}$, multiplied by the rate at which the particles meet, $R_\text{diff}$:
\begin{align}
 R_\text{LH} &= P_\text{react}\;R_\text{diff} \\
 &= P_\text{react}\frac{k_\text{diff,H$_2$}+k_\text{diff,OH}}{N_\text{sites}}n(\text{H}_2)n(\text{OH})
\end{align}
The overall process is bimolecular, of course. It depends on the surface concentrations $n(\text{H}_2)$ and $n(\text{OH})$, as well as on the concentration of adsorption sites $N_\text{sites}$.

When surface diffusion and microscopic sites are not explicitly included in the model, such as is commonly the case for two-, three- or multiphase rate-equation models, the competition between reaction, diffusion, and desorption of the reactants after they reside next to each other has to be taken into account. This can be done by calculating the probability to react as the ratio between the rate constant for reaction and the total rate constant for all processes:
\begin{equation}
 P_\text{react} = \frac{k_\text{react}}{k_\text{react}+k_\text{diff,H$_2$}+k_\text{diff,OH}+k_\text{des,H$_2$}+k_\text{des,OH}} \;. \label{rxn:h3o:competition}
\end{equation}

With this formulation of a LH rate, two limiting cases can be discussed. In
both, we will assume for simplicity that the desporption is negligible
compared to reaction and diffusion. The first case is a diffusion-limited
reaction, in which the diffusion of both species is much slower than the
reaction ($k_\text{diff}\ll k_\text{react}$). Then
$k_\text{diff}+k_\text{react}\approx k_\text{react}$ and
$P_\text{react}\approx 1$. Thus,
\begin{align}
  R_\text{LH,diffusion-limit} =
  \frac{k_\text{diff,H$_2$}+k_\text{diff,OH}}{N_\text{sites}}
  n(\text{H}_2)n(\text{OH}).
\end{align}
The other limiting case is a reaction-limited process in which the reaction
is much slower than the diffusion of both reactants, \emph{i.e.,} $k_\text{react} \ll
k_\text{diff,H$_2$}+k_\text{diff,OH}$. Then, $P_\text{react}\approx
k_\text{react}/(k_\text{diff,H$_2$}+k_\text{diff,OH})$ and 
\begin{align}
  R_\text{LH,reaction-limit} =
  \frac{k_\text{react}}{N_\text{sites}}
  n(\text{H}_2)n(\text{OH}).
\end{align}
Reaction rate constants, like the ones calculated in this work, influence the
overall Langmuir--Hinshelwood rate in the general case and in the
reaction-limited case. Since H$_2$ is assumed to diffuse fast, this is likely
to be the case for the reaction discussed in this paper.

Usually, for diffusion and desorption the approximation for the rate constant
\begin{equation}
  k_\text{process} = \nu_\text{trial}\;e^{-E_\text{process}/k_\text{B}T}
  \label{eq:trial}
\end{equation}
is made, where $\nu_\text{trial}$ is the trial frequency and $E_\text{process}$ the activation energy for diffusion or desorption. 

The reaction rate constant $k_\text{react}$ is provided by our instanton calculations. In models, however, often two approaches are tried and the rate constant that is the highest is chosen to be used in the model run: (a) the rate constant is calculated classically, analogous to \eqref{eq:trial} substituting $E_\text{process}$ with the reaction barrier and (b) tunneling is taken into account via a semiclassical approximation to the rectangular barrier approximation
\begin{equation}
k_\text{react} = \nu_\text{trial}\;e^{-2a/\hbar\;\sqrt{2\;\mu\;E_\text{reaction}}} \;.
\end{equation}
Here, $a$ is seen as the barrier width, but in fact can not be directly linked to any physical 
observable and $\mu$ is an effective  mass.
Another way to take into account tunneling is the use of the Eckart barrier approximation instead. 

We want to conclude  with a specific comparison between values for 
the reaction rate constants and kinetic isotope effect  calculated with the rectangular barrier approximation, the Eckart approximation (all with $\nu_\text{trial}$ kept constant), and instanton theory, taking into account the values published by \citeauthor{fur13}\cite{fur13} and \citeauthor{taq13}\cite{taq13}. 
The values for the rate constants and KIE are given in Table~\ref{tab:h3o:kie}. 
Firstly, it is clear that the choice for the value of the reduced mass has a strong influence on the rate constants and therefore on the KIEs, too. 
Furthermore, the rectangular barrier approximation is very crude and can underestimate the rate constants by several orders of magnitude.
This in turn can lead to wrong predictions of isotope fractionation in the ISM. 
Moreover, the rate constants themselves also span a large range between the different approaches and parameter choices which can have an effect on the thickness of the ice, the competition of OH reactivity with other species, and on the main route leading to water formation in dense molecular clouds. 
The Eckart barrier approximation works reasonably for this reaction although 
the reaction between H and H$_2$O$_2$ shows that an order of magnitude difference can appear between rate constants calculated with the Eckart approximation and the instanton method.\cite{lam16}
Also the KIEs calculated in this way give surprisingly good agreement with the KIEs obtained with the instanton theory.  

With this in mind, we wish to stress that it is important to realize when standard choices of parameter settings such as barrier width and reduced mass may not be enough to describe a reaction properly. In the case where better approximations are available, such as our instanton calculations, $k_\text{react}$ in \eqref{rxn:h3o:competition} can directly be taken as the unimolecular rate constant or the fit thereof (see Supplementary Information).

\section{Conclusions}

In this study, we computed chemical reaction rate constants of the reaction of hydroxyl radicals with molecular hydrogen (reaction~\ref{rxn:h3o:h2+oh}) on an $I_h$ ice surface. For this purpose, we used instanton theory on highly accurate potential energy surfaces. We provide reaction rate constants from 275\,K down to 110\,K for the ER mechanism (bimolecular) and down to 60\,K for the LH mechanism (unimolecular).
For both mechanisms, a fit of parameters of a modified Arrhenius equation was performed to obtain a continuous expression of $k(T)$.

To summarize the most important results:
\begin{itemize}
\item
For the reaction of H$_2$ and OH radicals, an ice surface just barely influences the potential 
energy along the reaction path, \emph{i.e.}, there is no classical catalytic effect.
Therefore, the surface effects can be included by using an implicit surface model.
\item
A surface can be implicitly mimicked by a structural gas-phase model, using the same rotational partition function for reactant and transition state. The reaction rate constants obtained in this way differ by 
a factor of 9.3 from the ones calculated on a full ice surface model.
\item
We found three different binding sites on our $I_h$ surface. The binding energy lies between 32 and 41 kJ mol$^{-1}$ (3850 and 4930\,K).
\item
The most important transition state structures and reaction paths are comparable to the ones in the gas phase. It follows that the adiabatic energy barriers (24--25\,kJ\,mol$^{-1}$) are similar to the barrier of the gas phase reaction (25.4 kJ mol$^{-1}$, 3055\,K).
\item
Kinetic isotope effects have been calculated for all possible isotope substitution patterns.
Exchanging the H to be transferred to D leads to a decrease in the rate constant 
of 2--3 orders of magnitude. Secondary KIEs are at most half an order of magnitude. 
\end{itemize}

\section*{Acknowledgments}
This work was financially supported by the German Research Foundation (DFG)
within the Cluster of Excellence in Simulation Technology (EXC 310/2) at the
University of Stuttgart and the European Union's Horizon 2020 research and
innovation programme (grant agreement No. 646717, TUNNELCHEM). T.L. wishes to
acknowledge the Alexander von Humboldt Foundation for generous support. The
authors acknowledge support for computer time by the state of
Baden-W\"urttemberg through bwHPC and the Germany Research Foundation (DFG)
through grant no INST 40/467-1 FUGG. Marie-Sophie Russ is thanked for
assembling the benchmark table and proofreading.  Thomas Bissinger is thanked
for the initial setup of the QM/MM calculations.

 \bibliography{mod}


\end{document}